\documentclass[12pt]{article}%
\usepackage{amsfonts}
\usepackage{amsmath}
\usepackage{amssymb}
\usepackage{graphicx}%
\providecommand{\U}[1]{\protect\rule{.1in}{.1in}}
\begin{document}

\title{IndIGO and LIGO-India: Scope and Plans for Gravitational Wave Research and
Precision Metrology in India}
\author{C. S. Unnikrishnan\thanks{On behalf of the LIGO Scientific Collaboration (LSC)
and the IndIGO Consortium. \ E-mail address: unni@tifr.res.in}\\Tata Institute of Fundamental Research, Mumbai 40005, India}
\date{Based on a talk at the Fifth ASTROD Symposium, Raman Research Institute,
Bangalore, July 2012. Close to the original published version LIGO document
P1200166. See www.gw-indigo.org for updates on LIGO-India.}
\maketitle

\begin{abstract}
Initiatives by the IndIGO (Indian Initiative in Gravitational Wave
Observations) Consortium during the past three years have materialized into
concrete plans and project opportunities for instrumentation and research
based on advanced interferometer detectors . With the LIGO-India opportunity,
this initiative has a taken a promising path towards significant participation
in gravitational wave (GW) astronomy and research, and in developing and
nurturing precision fabrication and measurement technologies in India. The
proposed LIGO-India detector will foster integrated development of frontier GW
research in India and will provide opportunity for substantial contributions
to global GW research and astronomy. Widespread interest and enthusiasm about
these developments in premier research and educational institutions in India
leads to the expectation that there will be a grand surge of activity in
precision metrology, instrumentation, data handling and computation etc. in
the context of LIGO-India. I discuss the scope of such research in the
backdrop of the current status of the IndIGO action plan and the LIGO-India project.

\end{abstract}

\section{Preamble}

The IndIGO Consortium for mobilizing and facilitating participation of Indian
scientists and engineers in gravitational wave (GW)\ observations was set up
in 2009 after a series of informal discussions during conferences and
workshops related to gravitation because both the need and fresh opportunities
for such participation was realized by the handful of people working in this
area for over two decades. It was felt that with the plans for advanced GW
detectors in the US and Europe, as well as with the possibility of new
projects in Australia and Japan, there was new opportunity opening up, missed
earlier due to limited resources during the era of initial GW interferometer
detectors like LIGO and Virgo. This turned out to be true when the idea was
strongly encouraged and supported by the global GW research community. For
over two decades, there have been important and influential contributions to
research in the theoretical aspects of the generation and detection of
gravitational waves from India (S. V. Dhurandhar at IUCCA, Pune on data
analysis strategies, methods and templates, and Bala Iyer at RRI, Bangalore on
GW generation processes, source modeling and waveforms). \ However, when the
small community of researchers in gravitation experiments and gravitational
waves discussed the possibility in the 1990s, the human and financial
resources that could be mustered in India were not perhaps anywhere near the
needs of the project and in any case no experimental activity in the field was
taken up at that time. Meanwhile, the community of researchers in those
theoretical aspects were steadily growing in India due to sustained research
at IUCAA and RRI as well as due to the possibility of signatures of primordial
gravitational waves in CMBR data attracting researchers in cosmology (like
Tarun\ Souradeep at IUCAA, Pune) to the task of detection of gravitational
waves. The possibility of an advanced large scale detector in India, after
suitable smaller steps of research and implementation, was discussed
enthusiastically again recently in the context of Advanced LIGO \cite{aLIGO}
and Virgo \cite{Virgo}, the upgraded GEO-HF \cite{GEO-hf} and the space-based
LISA \cite{LISA}, prompted and encouraged by several researchers (Bernard
Schutz, Rana Adhikari and David Blair). The formation of the IndIGO consortium
was followed up with a road-map for the Indian GW observational program that
was charted out during a meeting I organized with Munawar Karim on
experimental gravitation (ISEG2009) in Kochi in January 2009 \cite{iseg2009}.
This document was prepared in association with ACIGA, the Australian
consortium, represented in the meeting by D. Blair. It was natural to try to
participate in the ACIGA plans for a new advanced detector (AIGO) in
Perth,\ Australia \cite{AIGO} and it was envisaged that a limited yet active
participation in hardware and human resources for an advanced interferometer
detector was feasible and practical as a first serious step. Also, it was
decided that a prototype detector in India was an essential element of this
effort, for training and research. The overwhelming feeling during these steps
was that the opportunity was `now, or never'.

To assess the feasibility of this participation quantitatively, the IndIGO
consortium members sought a grant for `Establishing Australia-India
collaboration in GW Astronomy' in 2009, mainly to facilitate meetings between
AIGO and IndIGO, and this was immediately funded by the Australia-India
Strategic Research Fund (AISRF) managed by the Department of Science
\&\ Technology (DST) in India. Research in gravitational waves was already
recognized as a thrust area in the vision document of the DST. This was very
important because it enabled timely participation of the IndIGO
representatives in meetings in Shanghai (October 2009) and later in Perth and
Gingin in Australia (February 2010)\ where the idea of LIGO-Australia was
presented. LIGO-USA assessed the important scientific advantages of moving one
of the 4-km arm-length Advanced LIGO (aLIGO) detectors from Hanford to a
remote location outside the continent, and made the offer of contributing
important hardware of the interferometer to the Australian detector project.
Australian scientists had the responsibility for the infrastructure, assembly,
commissioning and operation of the detector. IndIGO envisaged a 15\%
participation in hardware deliverables and human resources and prepared a
project proposal to the DST seeking support. With LIGO-Australia
\cite{LIGO-Aus-DPR} as the primary project for our participation, it was
decided to attempt to construct a relatively small, yet advanced, prototype
interferometer at the Tata Institute of Fundamental Research (TIFR), Mumbai
where the availability of space is difficult, but facilities and support for
fabrication and testing etc. are excellent. The proposal for the 3-m scale
power recycled Michelson-Fabry-Perot interferometer with a conservative budget
of about \$500,000 went through extensive reviews, being a new endeavour, and
was successful in getting funded within a year, in 2011. Several aspects of
Indian collaboration for LIGO-Australia was discussed in the India-Australia
meeting under the AISRF grant, during February 2011 in Delhi. Based on a joint
proposal with IUCAA, Pune and Caltech, USA as the node institutions, IndIGO
obtained a grant for an Indo-US centre for gravitational wave physics and
astronomy from the Indo-US Science \&\ technology Forum (IUSSTF) in July 2011,
to facilitate mutual visits and joint work. Another significant development
was that the IndIGO consortium became a member of the Gravitational Wave
International Committee (GWIC), in July 2011. However, there were serious
difficulties for substantial funding in Australia for realizing the large
scale LIGO-Australia detector within a reasonable schedule, set as about mid-2011.

The flavour of the plans for GW research in India changed drastically after
mid-2011 with the dawn of the possibility of a LIGO-India detector, to be
constructed and operated in India with the hardware for the interferometer
from LIGO-USA and the infrastructure to house it provided by India. The
arrangement for LIGO-India is similar to what was envisaged for the
LIGO-Australia detector, with the additional, and very important, task of
selecting a suitable site in India for locating the detector. This `amazing
opportunity', as I call it, was facilitated due to both the mutual trust and
confidence developed during several interactions between the US and Indian
scientists in GW research as well as the support for the idea from key
researchers in the field of gravity, like A. Ashtekar, K. Thorne, and B.
Schutz. An important milestone in this fast-paced developments was the IndIGO
consortium becoming a member of the LIGO Scientific Collaboration (LSC) in
September 2011. The plan for LIGO-India was well in accordance with the goals
of development of the field, especially for international collaboration and
network operation, outlined in the road-map of the GWIC \cite{GWIC-rmap}. In
October 2011, a panel of the National Science Foundation (NSF), USA reviewed
the case LIGO-India and found its science case compelling enough to go
forward, albeit with cautious evaluation criteria. Ever since, the LIGO-USA
team has been working relentlessly to ensure with proper evaluation that the
LIGO-India detector is feasible, by setting and examining several target
criteria. The scientists, science mangers and the funding agencies in India
were already sensitized to the importance of the national participation in GW
research and astronomy, from the several meetings and discussions in the
context of IndIGO and LIGO-Australia project, and this enabled IndIGO to
prepare a detailed project proposal \cite{LIGO-India}, which was submitted to
the potential funding agencies - the Department of Science and Technology and
Department of Atomic Energy, Government of India. This was discussed in a
meeting in November 2011 and LIGO-India proposal received enthusiastic support
and encouragement along with several other astronomy mega-projects. This paved
way for the Planning commission of India discussing the project as a potential
`12th plan' mega-science project, to be initiated during the 12th 5-year plan
of the government of India, during 2012-2017. When the Inter-University Centre
for Astronomy and Astrophysics (IUCAA), Pune, and the two key technologically
highly endowed institutes under the DAE -- The Institute for Plasma Research
(IPR), Gandhinagar and the Raja Ramanna Centre for Advanced Technology
(RRCAT), Indore -- agreed to take key responsibilities for the projects,
things fell into sharp focus. Speedy and cautious response and support from
the NSF, USA in the form of visits of key persons and reviews of the proposal
by special committees resulted in quickly giving concrete form to the
LIGO-India project. Four senior level visits from the LIGO-Laboratory to the
LIGO-India lead-institutions for technical assessment and discussions were
followed by three in depth reviews by a NSF panel. All this culminated in a
review by the National Science Board, USA, in August 2012 and the following
resolution: `Resolved, that the National Science Board authorize the Deputy
Director at her discretion to approve the proposed Advanced LIGO Project
change in scope, enabling plans for the relocation of an advanced detector to
India'. The Department of Atomic Energy is putting together the papers for a
note to the Cabinet seeking "in-principle" approval of the project, and
permissions to sign the relevant MOUs and release of seed funding for the
project. It is significant that LIGO-India is listed prominently in the
communication from the US state department on U.S.- India Bilateral
Cooperation on Science and Technology \cite{US-state-dept}.

\section{The LIGO-India project: Science case}

The idea of LIGO-India (as well as the earlier LIGO-Australia) arose due to
compelling science reasons when it was realized that a network of three
advanced detectors, preferably of the same nature of design and sensitivity,
forming a large triangle across the globe has significantly more advantages
than the operation of the same three detectors in just two geographical sites,
as was envisaged in the LIGO plan. In addition such a change of plan has the
advantage of bringing in a new country and scientific community into the
global GW research and astronomy effort. Of course, there is a change in the
noise cancellation capability at one of the sites where the two detectors were
supposed to operate simultaneously in the same vacuum enclosure (Hanford in
this case). However, that was seen as a small price to pay for the great
advantages LIGO-India would offer in terms of source localization, duty cycle,
sensitivity and sky coverage \cite{Klimenko,Fairhurst,Fair-Sathya}. A study by
B. Schutz, in which a general framework based on three new figures of merit
was developed for studying the effectiveness of networks of interferometric
gravitational wave detectors, showed that enlarging the existing LIGO--Virgo
network with the planned detectors in India (LIGO-India) \ and Japan (LCGT)
brought major benefits, including much larger detection rates and more uniform
antenna pattern and sky coverage \cite{Schutz2011}. I summarize here the main
scientific advantages that are discussed in these papers.

\subsection{Source localization}

The obvious advantage of locating a third aLIGO detector far away from the
other two in the USA is the ability to locate a source in the sky with three
identical detectors, with an accuracy of about a degree or so. In the
pre-LIGO-India plans this was expected to be accomplished by timing
measurements involving the aLIGO detectors in two locations and the Virgo
detector at Cascina near Pisa. This of course is possible. However, there is
significant improvement in the source localization ability when the LIGO-India
detector is added to the network. The advantage is quantitatively similar with
the third aLIGO detector in either Australia or India due to the bounded
positioning possibilities on the spherical earth, with marginal advantage
(10\%) in the slightly larger baselines to Australia from the US sites (there
is significant difference to the baseline to Virgo, however, of about 40\%).
Figs. 1 and 2, from ref. \cite{Fair-Sathya}, indicate the locations of the
detectors and the improvement in localization. (H - Hanford, USA, L -
Livingston, USA, V - Virgo, Pisa, Italy and I - LIGO-India, assumed to be
located near Bangalore for these estimates. The sensitivity of localization
error to the exact position in India is mild). Since binary neutron star
mergers are the most promising events for detection when the advanced
detectors start their operation, the analysis in this paper is based on a
population of binary neutron stars distributed over the sky at a luminosity
distance of 200 Mpc, corresponding to the average distance reach of the
advanced detectors.%

%TCIMACRO{\FRAME{ftbpFU}{2.6693in}{2.6782in}{0pt}{\Qcb{Schematic detector
%locations with maximum delays in milliseconds. H: Hanford, L: Livingston, V:
%Virgo and I: LIGO-India. (Figure credit: Ref. \cite{Fair-Sathya}).}}%
%{}{fig1.eps}{\special{ language "Scientific Word";  type "GRAPHIC";
%width 2.6693in;  height 2.6782in;  depth 0pt;  original-width 4.5938in;
%original-height 4.6099in;  cropleft "0";  croptop "1";  cropright "1";
%cropbottom "0";  filename 'Fig-new/fig1.eps';file-properties "XNPEU";}} }%
%BeginExpansion
\begin{figure}[ptb]%
\centering
\includegraphics[
height=2.6782in,
width=2.6693in
]%
{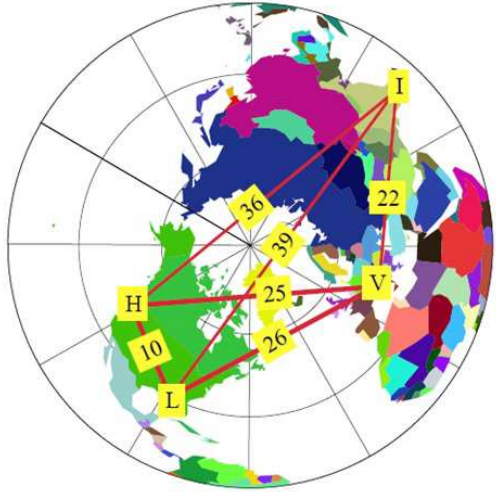}%
\caption{Schematic detector locations with maximum delays in milliseconds. H:
Hanford, L: Livingston, V: Virgo and I: LIGO-India. (Figure credit: Ref.
\cite{Fair-Sathya}).}%
\end{figure}
%EndExpansion
%

%TCIMACRO{\FRAME{ftbpFU}{4.9617in}{1.3865in}{0pt}{\Qcb{The remarkable
%improvement in the source localization error is evident in the comparison of
%the figures for HLV and HLVI. (Figure adapted from Ref.
%\cite{Fairhurst,Fair-Sathya}).}}{}{fig2.eps}%
%width 4.9617in;  height 1.3865in;  depth 0pt;  original-width 10.0495in;
%original-height 2.7838in;  cropleft "0";  croptop "1";  cropright "1";
%cropbottom "0";  filename 'Fig-new/fig2.eps';file-properties "XNPEU";}} }%
%BeginExpansion
\begin{figure}[ptb]%
\centering
\includegraphics[
height=1.3865in,
width=4.9617in
]%
{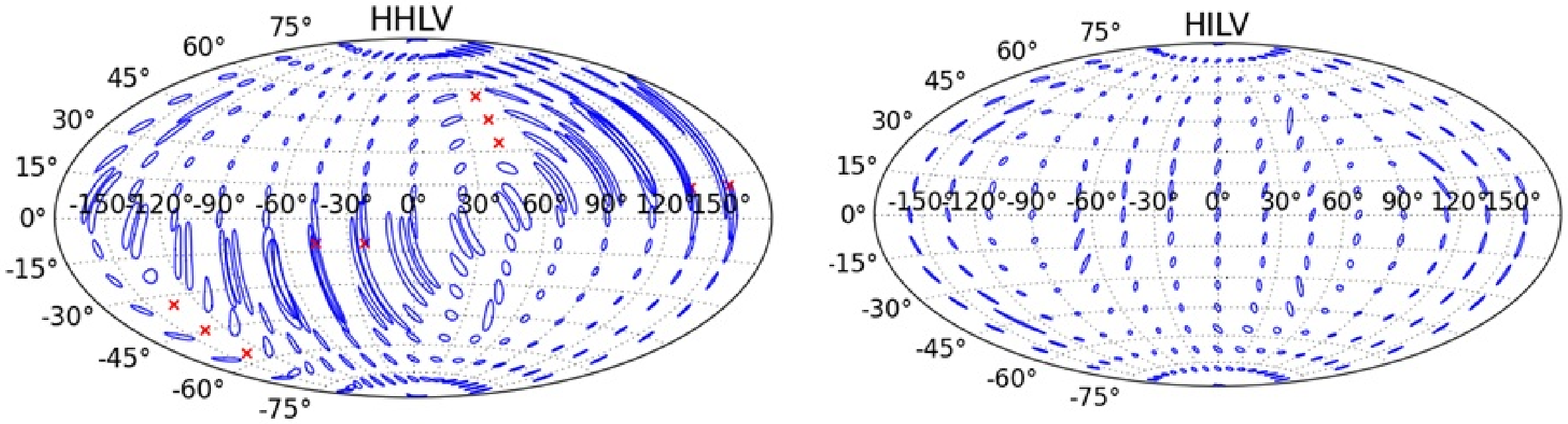}%
\caption{The remarkable improvement in the source localization error is
evident in the comparison of the figures for HLV and HLVI. (Figure adapted
from Ref. \cite{Fairhurst,Fair-Sathya}).}%
\end{figure}
%EndExpansion

With a baseline of about 14000 km, the theoretical resolution for localization
is about 10 sq. degree. Of course the details depend on the projection of this
baseline on the sky, and hence varies with direction. With the four-detector
network involving LIGO-India, HLVI, there is dramatic improvement, by an order
of magnitude in the worst case of just the HHLV network. For 50\% of the
sources the average localization error improves from 30 sq. deg to 8 sq. deg.
Apart from localizing the source on the two-dimensional sky, it is also
possible to measure the distance to the source because binary neutron stars
are standard candles, and it is estimated that four or three-detector networks
involving LIGO-India can measure distances to better than 30\%, the
improvement resulting from the large baseline as well as the ability to sense
the orbital orientation of the binary relative to the line of sight. These
measurements will be important for the independent determination of the Hubble parameter.

\subsection{Sensitivity}

There will be a 10-fold improvement in the sensitivity of individual Advanced
LIGO and Virgo detectors, compared to the earlier versions. In network
operations, the overall sensitivity is determined by noise rejection
capabilities achieved in coincidence detection. Therefore, the original HHL
configuration had extra sensitivity coming from the fact that there are two
Hanford detectors in the same UHV enclosure. Together with the Livingston
detector, the sensitivity for detection along the normal to the US continent
was almost twice or even 3 times of that in other direction rotated 90 degrees
away (see left panel of Fig. 3). The sensitivity is smoothed out more
uniformly in all directions in the HLVI configuration \cite{Klimenko}. Even
though there is marginal reduction in the best sensitivity (about 15\%), the
worst sensitivity in some directions is only about half of that of the best
sensitivity and significant region of the sky is visible with much better
sensitivity (20\%-30\%) than possible with the HHLV configuration.%

%TCIMACRO{\FRAME{ftbpFU}{5.3179in}{1.5602in}{0pt}{\Qcb{Comparison of
%sensitivity for detection in the original HHLV configuration and the proposed
%HLVI configuration. While it is true that the sensitivity diminshes around the
%normal vector of the HL detector locations (and 180 degree across), there is
%signficant improvement in other regions (reduction of blue patches). (Plot
%credit: P. Ajith, IndIGO).}}{}{fig3.eps}%
%width 5.3179in;  height 1.5602in;  depth 0pt;  original-width 7.5172in;
%original-height 2.1841in;  cropleft "0";  croptop "1";  cropright "1";
%cropbottom "0";  filename 'Fig-new/fig3.eps';file-properties "XNPEU";}} }%
%BeginExpansion
\begin{figure}[ptb]%
\centering
\includegraphics[
height=1.5602in,
width=5.3179in
]%
{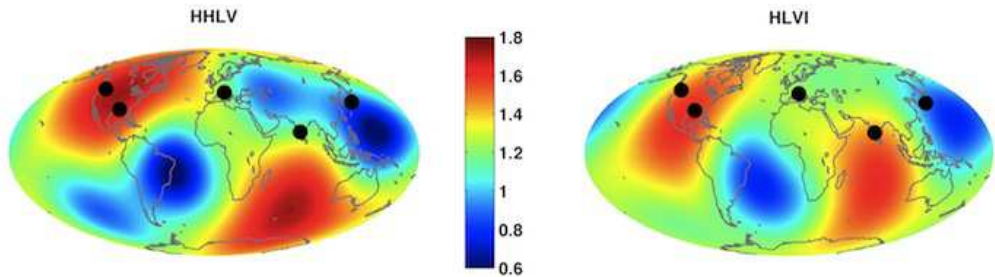}%
\caption{Comparison of sensitivity for detection in the original HHLV
configuration and the proposed HLVI configuration. While it is true that the
sensitivity diminshes around the normal vector of the HL detector locations
(and 180 degree across), there is signficant improvement in other regions
(reduction of blue patches). (Plot credit: P. Ajith, IndIGO).}%
\end{figure}
%EndExpansion

\subsection{Sky coverage}

Due to the fact that the new detector is significantly out of plane compared
to the HLV configuration, the blind bands in the HHLV configuration are
eliminated considerably in the HLVI configuration. Compared to a sky coverage
of about 47\% in the case HHLV, the HLVI network has 79\% sky coverage. With
the addition of the KAGRA detector in Japan (HLVIJ) this reaches 100\%,
compared to 74\% of HHLVJ. The details are listed and discussed in ref.
\cite{Fair-Sathya}.

\subsection{Duty cycle}

Each advanced detector is expected to have a duty cycle below 80\%, due to the
need for regular maintenance. Regular operation with good precision in
localization requires a minimum of three detectors and therefore the duty
cycle for a 3-detector operation is limited to about 52\% in the HHLV
configuration where the down time of one H detector has significant overlap
with the down time for the other because they share the same UHV envelop,
similar infrastructure etc. However, with HLVI, there are four 3-detector
configurations available with total effective duty cycle of 41\%, equal to the
four-detector duty cycle itself, adding up to 82\%. This is a significant
improvement for the `on' time for the network telescope, and this fact alone
is worth relocating one detector to India, from the point of view of astronomy
(Fig. 4). Of course, source localization is still considered as the most
important single factor in the many scientific advantages of LIGO-India
because it is a crucial factor in source identification with simultaneous
observations with other types of telescopes operating in the electromagnetic spectrum.%

%TCIMACRO{\FRAME{ftbpFU}{3.6064in}{2.2638in}{0pt}{\Qcb{Improvement in the duty
%cycle with the 3-detector operation. (Figure credit: Ref. \cite{Fair-Sathya}%
%).}}{}{fig4.eps}{\special{ language "Scientific Word";  type "GRAPHIC";
%width 3.6064in;  height 2.2638in;  depth 0pt;  original-width 5.6348in;
%original-height 3.5286in;  cropleft "0";  croptop "1";  cropright "1";
%cropbottom "0";  filename 'Fig-new/fig4.eps';file-properties "XNPEU";}} }%
%BeginExpansion
\begin{figure}[ptb]%
\centering
\includegraphics[
height=2.2638in,
width=3.6064in
]%
{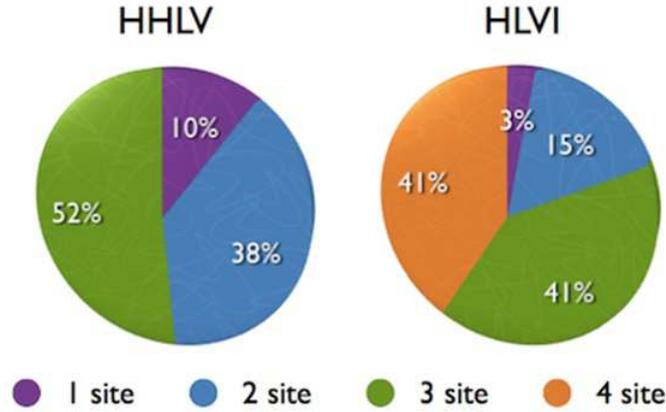}%
\caption{Improvement in the duty cycle with the 3-detector operation. (Figure
credit: Ref. \cite{Fair-Sathya}).}%
\end{figure}
%EndExpansion

\section{The LIGO-India project: Plan of execution}

The LIGO-India project \cite{LIGO-India} is a joint endeavour between the
consortium of research laboratories and universities funded by the Government
of India and the LIGO Laboratory of the USA, funded by the National Science
Foundation, USA. The LIGO Laboratory will provide the complete design and all
the key detector components of the Advanced LIGO (aLIGO) detector. These
include the vibration isolation platforms and systems, the pre-stabilized
laser with amplifier, complete optics and suspension systems for mode cleaners
and the interferometer, sensors, control systems and electronics, software,
design and \ assembly drawings and documents etc. India would provide the
infrastructure and human resources to install the detector at a suitable site
in India and would be responsible for commissioning and operating it. The
infrastructure involves the appropriate site, suitably prepared, UHV
enclosures and the 2x4km beam tubes, laboratories and clean rooms etc. The aim
is to realize a third aLIGO detector, as close as possible in design and
operating characteristics to the other two aLIGO detectors in the USA, in time
to enable GW astronomical observations in a multi-detector network. The
proposed observatory would be operated jointly by IndIGO and the LIGO
Laboratory and would form a single network along with the LIGO detectors in
USA and Virgo in Italy, the advanced detectors under assembly and
commissioning, and possibly the KAGRA detector in Japan, which is in its
beginning stage of construction. It will bring together scientists and
engineers from different fields like optics, lasers, gravitational physics,
astronomy and astrophysics, cosmology, computational science, mathematics and
various branches of engineering. In order to fully realize the potential of
multi-messenger astronomy, the LIGO-India project will join forces with
several Indian astronomy projects. Potential collaborators include the
space-based `Astrosat' multi-wavelength astronomy project \cite{Astrosat}, the
high altitude gamma-ray observatory (HAGAR)\cite{hagar}, the India-based
Neutrino Observatory (INO) \cite{INO}, the Giant Meter-wave Radio Telescope
(GMRT) \cite{GMRT} and other optical/radio telescopes.

The total estimated budget over 15 years is about \$250 million on the Indian
side and the value of the interferometer hardware from LIGO-USA, including the
research input, is estimated at about \$120 million. The arrangement
drastically cuts down the time required, perhaps by 8-10 years, to fabricate
the key components and modules to specifications, assemble and operate a new
advanced detector at a third location with large baseline, and this is of
course the greatest advantage from the point of view of Indian scientists. By
using exactly the same hardware and software as the aLIGO detectors, the
confidence in the quality and understanding of the data collected in the 3
similar detectors is ensured to large extent. The two aLIGO detectors are now
operational in their first observational run (O1) with about 1/3 of their
projected sensitivity. The assembly and testing of the LIGO-India detector is
expected to start in 2019, and by then the two aLIGO detectors and the Virgo
detector will be fully operational in the network and it is likely that the
KAGRA detector will be ready as well. This staggered schedule, inevitable due
to the large infrastructure to be prepared, especially the UHV enclosures and
beam tubes, also helps in eventual speed up during 2018-22. Some of these
details are discussed in the LIGO-India `detailed project report' (DPR) public
document, ref. \cite{LIGO-India}. The reasonable expectation from this tight
schedule requiring systematic and sustained work is that India will operate an
advanced gravitational wave detector with strain sensitivity similar to the
aLIGO and advanced Virgo detectors by 2022.

\section{The LIGO-India project: National participation}

A long term megascience project like the LIGO-India can be taken up and
completed only with an inclusive national participation involving research
institutes with different expertise and experience. In addition, strong
industry participation is necessary while creating the infrastructure,
especially the UHV enclosures and the clean room environments. Fortunately,
IndIGO has been able to identify and mobilize several key research institutes
in India to take up the challenge and contribute in various ways to the
project. Most importantly, three major institutes volunteered to play the lead
role, with crucial contributions of deliverables in their areas of expertise.
These are the Institute for Plasma Research (IPR), Gandhinagar (near
Ahmedabad), Raja Ramanna Centre for Advanced Technology (RRCAT), Indore and
the Inter-University Centre for Astronomy and Astrophysics (IUCAA), Pune. The
IndIGO consortium that started out with several theoretical and data analysis
scientists and two or three experimental physicists has now grown to a large
membership of over 120, with half its strength coming from experimenters and
engineers. The possibility of the mega-science detector project `at home' with
a global presence and collaboration is unprecedented and it is certain that a
larger community will form as we progress through the detector construction
towards operation.

\subsection{University participation}

In India, fundamental science projects that involve large scale infrastructure
and collaboration of a large number of people have been pioneered and managed
by some of the national laboratories, notably the Tata Institute of
Fundamental Research, Mumbai. Indian universities, by and large, have kept
away from large scale national projects due to constraints of funding and
other administrative difficulties, even though small groups have been
participating as partners in accelerator based particle physics research for
some time now. The Indian Institutes of Technology (IIT) faculty have been
very active in several projects of small and large scales with a technology
flavour, but have been minor partners in large scale astronomy or physics
projects. The Inter-University Centres were set up to increase the
participation of university scientists in fundamental research through
centralized facilities. The LIGO-India project has generated tremendous
enthusiasm among both communities and the timing coincides with the new
national initiative for high quality undergraduate education through several
newly set up institutes (IISER) in several states of India. We expect that the
LIGO-India project will bring together researchers and students from
universities, IITs and IISERs in an unprecedented scale and level of integration.

\subsection{Industry participation}

From infrastructure development in the initial stages to the commissioning of
the detector, strong industry participation is required for the success of
LIGO-India project.\texttt{ }The large scale sophisticated infrastructure
required to house the GW detector, involving technologies for UHV, steel
processing, robotic welding, clean rooms, hydraulics, computing clusters,
power handling etc. can be realized only with the participation of the
relevant industries. Particularly important is the creation of ultra-clean
laboratory environment and ultra-clean ultra-high vacuum. Therefore, ensuring
participation of some of these industries, especially those related to UHV
technology, in some of the discussion meetings leading to the LIGO-India
project has been a priority. Even though some of the fabrication aspects that
are LIGO-specific are new for the industrial partners that we have started to
identify, there is a sense of new national adventure in these circles that
will help them to take up the challenges. This is possible only by adhering
well to the schedule and budget expenditure plan. The lead institutes and the
IndIGO consortium council are keenly following up the specific needs of the
LIGO-India project in this context with the funding agencies and policy makers.

We will be interacting strongly also with the electronics and computer
industrial sources and a mutually beneficial long term partnership is
envisaged. When the detector is operational a large amount of computing in the
cloud environment is expected and it is natural that the project will
contribute to the development in distributed computing in India.

The detector components, contributed by LIGO-USA define several goals for
technology achievements within the country for optical components, sensor
technologies, feed-back control systems, mechanical fabrication etc. Our
interaction with the industry in the context of ideas and hardware for next
generation detectors will help to realize some of these goals with benefits of
global exposure and market for relevant specialized industries.

\subsection{Outreach and higher education}

There are very few internationally competitive physics and astronomy
instruments operating within India. The lack of accessible and visible
facilities for research that involves high technology and high finesse
instrumentation directly affects the interest and motivation for choosing
experimental physics and engineering physics as a career, especially in areas
that requires a long term commitment and field work. The LIGO-India project
will dramatically change this situation and a large number of tested remedial
steps will built into the operation of the detector for continued engagement
with undergraduate and post-graduate students in physics, astronomy and
engineering. This will include introductory and advanced schools, hands-on
laboratories at site as well as in associate centres at universities, IISERs
and IITs, summer training programs at associated laboratories in India and
abroad for selected motivated students, special LIGO-India fellowships for
specialized training etc. The LIGO-India detector will be one of the very few
research facilities in India of this scale, international relevance and
technological innovation to which the general public and students can have
access through an interface centre located not far from to the actual
detector. It has the additional fascination as an instrument for astronomy of
the neutron stars and black holes in distant galaxies. Creation and operation
of a public outreach centre where key technologies and physical principle that
make the detector will be on display, some of which for hands-on access, is an
integral part of the project, as has been the practise in LIGO-USA, AIGO etc.
as well. This will also serve, through direct interaction and through
web-based services, as a centre for continuing education on these topics.
Subsidiary centres of a similar nature will also be set up in the associate
centres of the IndIGO consortium for wider reach, especially to school
students. We will also tie up with national planetariums in different cities
in India for programs on gravitational wave astronomy and teacher training in
related areas. The home computer based 'Einstein@home' program for data
analysis for continuous wave sources that is currently running as part of the
gravitational wave detection global activity will add to the public outreach
program. It is not inconceivable that with the enthusiasm among students in
India to get involved in such projects this program will touch a 1 million
user mark and 1000 Tflops peak.

\subsection{Synergy of funding sources}

A national megaproject of this scale and spread of participation needs
synergic funding and support from multiple national agencies. Indeed, the
project proposal was considered by the two prominent national agencies that
support fundamental research in physics and astronomy in India - the
Department of Science and Technology (DST) and the Department of Atomic Energy
(DAE). Projects requiring large financial support need to generate a consensus
among scientists, science planners and public funding agencies. The IndIGO
consortium prepared and submitted the LIGO-India proposal in time before the
National Planning Commission finalized its allocations for the fresh 5-year
plan for revenue spending in India (2012-17), after making several
presentations in several forums, stressing the imminent detection of
gravitational wave and the potential and importance of ensuing GW astronomy.
In a joint meeting of representatives of the entire astronomy community in
India, which discussed several large scale astronomy projects, LIGO-India
received enthusiastic support, and the two agencies (DST\ and DAE) agreed to
include the LIGO-India proposal in the list of Mega-Projects being considered
by the planning commission of India.

\section{Prototype detector at TIFR, Mumbai}

It was realised from the outset that an advanced prototype interferometer
detector that incorporates all essential features of the large scale detector
is an important element in the road-map for GW research and astronomy in
India. Therefore a detailed proposal for 3-m scale power-recycled
interferometer detector was prepared in 2010 and submitted to the TIFR, Mumbai
where associated expertise and facilities for taking up such a project
existed. After extensive discussion and evaluation this project at an
estimated cost of \$500,000 was approved. However, the need to construct an
entirely new laboratory building and the delay in fund flow as well as in the
approvals for LIGO-India have introduced some schedule uncertainty. It is
expected that there will be an operational interferometer with a displacement
sensitivity of about $10^{-17}m/\sqrt{Hz}$ by 2018, in the newly constructed
laboratory. I now summarize the main design features and goals of this
prototype detector (Fig. 5).%

%TCIMACRO{\FRAME{ftbpFU}{4.5356in}{3.4203in}{0pt}{\Qcb{Feature summary of the
%small prototype detector.}}{}{fig5.eps}{\special{ language "Scientific Word";
%type "GRAPHIC";  maintain-aspect-ratio TRUE;  display "USEDEF";
%valid_file "F";  width 4.5356in;  height 3.4203in;  depth 0pt;
%original-width 7.1287in;  original-height 5.3671in;  cropleft "0";
%filename 'Fig-new/Fig5.eps';file-properties "XNPEU";}} }%
%BeginExpansion
\begin{figure}[ptb]%
\centering
\includegraphics[
height=3.4203in,
width=4.5356in
]%
{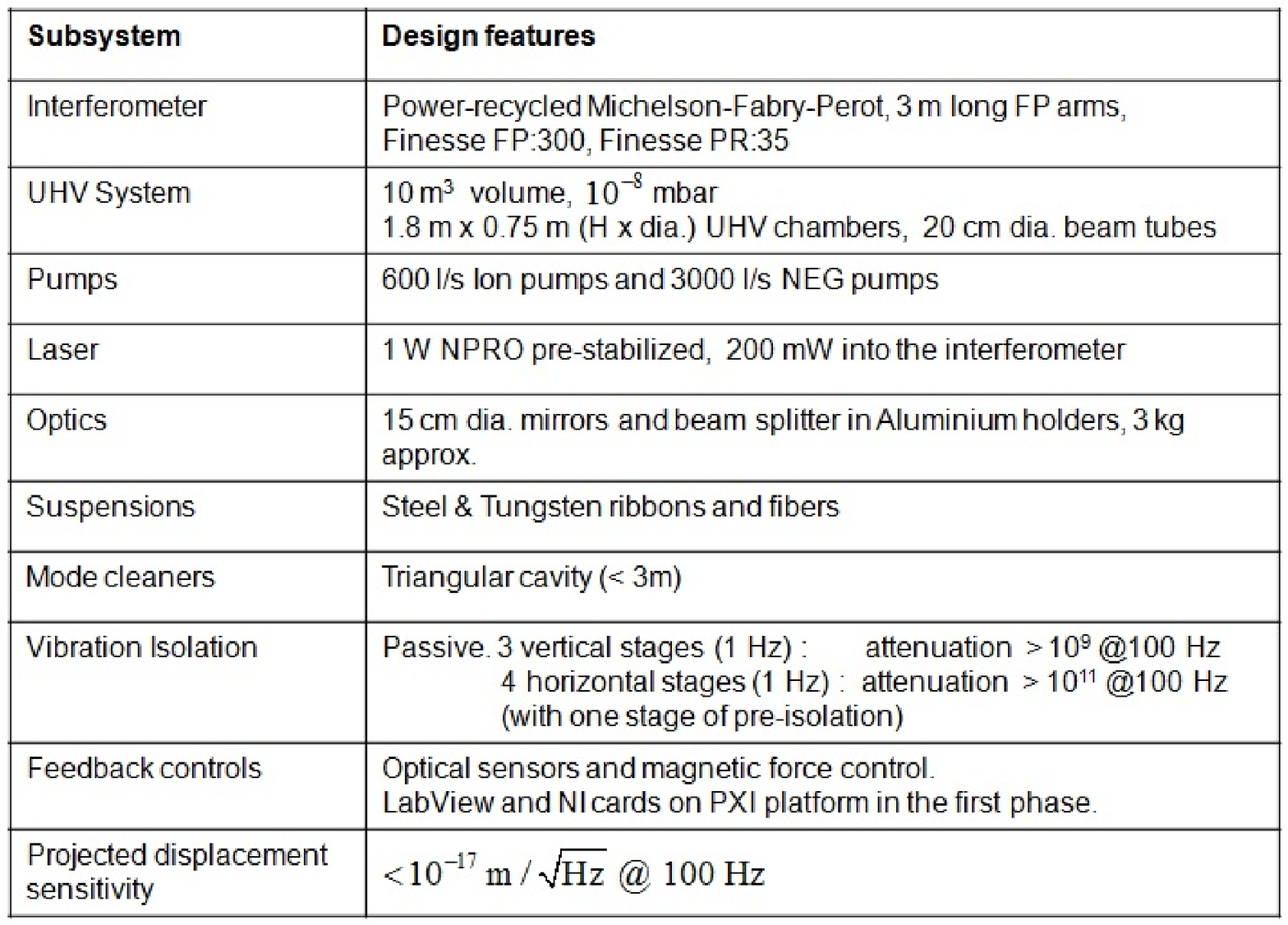}%
\caption{Feature summary of the small prototype detector.}%
\end{figure}
%EndExpansion

The prototype detector is expected to serve as the research and training
platform with all the features of the aLIGO-like detectors, scaled down to
displacement sensitivity around $10^{-17}m/\sqrt{Hz}$ above 100 Hz. It will be
the Indian research platform for features like signal recycling, DC read-out,
and most importantly the use of squeezed light and noise reduction for
precision metrology. It is envisaged that some parallel development on
squeezed light based measurement technologies will be developed and this will
be implemented in the prototype interferometer after 2018. It will also serve
as a superb instrument for novel studies on short range gravity and QED force,
especially a measurement of the Casimir force in the range 10 -100 microns
where no previous measurements exist \cite{Raji-unni}. The main idea here is
that even though the response of the suspended end mirror to a modulated force
at frequency $\omega$ contains the attenuation factor $\left(  \omega
_{0}/\omega\right)  ^{2},$ the already fine displacement sensitivity
$10^{-17}m$ can be enhanced with integration to below $10^{-19}m$ over several
hours. This makes the Casimir force measurable with good precision for
separations larger than 10 microns. Possible coincidence operation with next
generation cryo-mechanical detectors by optimizing the sensitivity at 2kHz+ by
signal recycling and use of squeezed light (at strain sensitivity approaching
$10^{-20}m/\sqrt{Hz}$) is also envisaged.

\section{Satellite projects and Precision metrology}

A natural outcome of a project like LIGO-India is its catalyzing ability. With
a large number of advanced technologies that have been stretched to their
present limits in use in LIGO-India, the project is at once a model system for
the use of similar technical strategies in other areas of precision metrology
of both fundamental and practical nature, and a motivating platform for the
next generation technologies. Only when the limits are visible, one is spurred
into conceiving the next generation of technologies. In our laboratory,
satellite projects to develop matter-wave interferometers based on both
ultra-cold atoms and liquid helium will be taken up with a view to contribute
to next generation gravitational experiments, potentially including GW
detection. Cold atom interferometers are just an additional step or two from a
\ wealth of technologies that we already developed at TIFR, working with
ultra-cold Rb and K atoms as well as Rb Bose-Einstein condensate. The fact
that the gravitational coupling energy and therefore the gravitationally
induced phase for an atom with a mass of 100 GeV is about $10^{11}$ times the
gravitational energy of a 1 eV photon is the major advantage of a matter-wave
interferometer exploring gravity. However, the size of matter-wave
interferometers are tiny compared to optical interferometers and the possible
configurations are limited. Yet, it is speculated that these might allow
breakthroughs in several types of gravitational experiments, including the
detection of gravitational waves \cite{Kasevich}. Matter wave interferometers
are also important in addressing foundational questions on the quantum
dynamics of particles in gravitational fields \cite{Unni-gill-PLA2012}.

It is expected that a significant number of researchers in IISERs and IITs
will take up small scale projects associated with LIGO-India aimed at
technologies to be incorporated in the next generation detectors. Novel sensor
modules and electronics, UHV compatible devices, compact and sensitive tilt
meters, displacement and angle sensors, accelerometers and gyroscopes,
computing strategies, integrated optics etc. are some examples. Due to the
stringent requirements on the spatial placements of LIGO optical components,
high precision survey has been an important aspect of the LIGO installation,
and the methods used in defining the coordinates for LIGO-India precision
installation are expected to contribute to geophysical measurements and survey
strategies. The LIGO-India detector itself, equipped with its elaborate active
vibration isolation system, is a sensitive instrument to monitor geophysical
phenomena in the acoustic frequency range.

\section{LIGO-India and the Indian space science program}

It is clear that the greatest potential for gravitational wave astronomy
resides in the wavebands in the range $10^{-5}$ Hz - $1$ Hz$.$ Only a
space-based detector with sufficient large arm length in that relatively
noise-free environment can sense these signals. Even though the original
LISA\ project is currently not realized, new thinking on this possibility will
remain active due to its importance, and the inevitability from the point of
view of the physicist and astronomer. Other possibilities are also being
discussed, like the ASTROD-GW (China) \cite{Astrod}, eLISA (ESA) \cite{eLISA}
or DECIGO (Japan) \cite{Decigo}. In addition there are a few space-based
gravitation experiments of high significance, like the GG \cite{GG}, which can
benefit from the involvement of an active space-based fundamental physics
program in India. We expect that LIGO-India will motivate new initiatives
within the highly successful Indian Space Research Organization itself towards
new astronomy and fundamental physics.

\section{LIGO-India: Schedule and Progress}

Though many of the steps in bringing the proposal to the final stages of
approvals have been taken, the final approval and funding allocation for the
project from the cabinet committee of the government of India is still
awaited, as of October 2015. Given this situation, there are constraints on
going ahead with the initial tasks on site identification and selection,
identifying potential partner institutions and industries, visits to LIGO-USA
etc. However, the lead institutes have been active on the project and have now
identified expert teams within, for several technical tasks of LIGO-India.
Working visits from IndIGO members as well as training of researchers,
engineers and post-doctoral fellows at the LIGO observatories at Hanford and
Livingston in the several technical aspects of the aLIGO detectors is an
intergral part of the early schedule. The EGO consortium in Europe, managing
the Virgo detector, also has extended help in the matter of training and
technology exposure. The IndIGO consortium members have been working on the
initial tasks with support from the lead institutes. Most significantly,
several potential sites in several states of India have been visited and
preliminary measurements on seismicity and environmental noise factors have
been conducted. The policy is not to compromise on the strict constraints laid
out for a suitable site in terms of its long term isolation from seismic and
man-made noise, while keeping in mind the accessibility for construction and
operation on schedule.

The ground noise models at the LIGO Hanford and Livingston sites are
indicated, for reference, along with the requirement on residual noise after
the pre-isolation stage, in figure 6. The active pre-isolation system for LIGO
is already designed and fabricated to bring down the ground vibration noise at
the two LIGO sites to the required level, \ with isolation factor of about
1000 in the frequency range 1-30 Hz.

India has several low seismicity areas, usually indicated as zone II in
seismic activity maps in India. From short term seismic surveys conducted over
two weeks each, it is now known that isolated areas in these zones are
seismically quiet enough to install an advanced gravitational wave detector.
Many potential sites that are more than 150 km from the ocean in such areas,
spread around India in the Deccan plateau in Karnataka and Andhra Pradesh, and
in Madhya Pradesh, Rajasthan and Chathisgarh, for example, have been visited
for preliminary evaluations. Long term weather data from the meteorology
department or from the repository at the nearest airport , as well as about 2
weeks of seismic noise data have been obtained. Typical ground noise measured
in such sites during quietness is shown in the figure 7, obtained from IndIGO
measurements spanning about 2 weeks with a 3-axis wideband Gurlap seismometer
(more recent extended measurements confirm this trend). Since the typical
ground noise at the sites we briefly explored in this frequency range seems
comparable to that in the LIGO sites, the same isolators are expected to
ensure their intended isolation levels in LIGO-India .%

%TCIMACRO{\FRAME{ftbpFU}{4.8319in}{1.9675in}{0pt}{\Qcb{Left: Sketch of aLIGO
%projected strain sensitivity. Right: Approximate displacement noise spectral
%densities at the two LIGO sites Hanford and Livingston, along with the
%requirement (solid curve) on displacement noise where the seismic isolation
%system supports the test mass suspension (from ref. \cite{Abbot--rana-CQG}).}%
%}{}{fig6.eps}{\special{ language "Scientific Word";  type "GRAPHIC";
%width 4.8319in;  height 1.9675in;  depth 0pt;  original-width 10.0343in;
%original-height 4.0647in;  cropleft "0";  croptop "1";  cropright "1";
%cropbottom "0";  filename 'Fig-new/Fig6.eps';file-properties "XNPEU";}} }%
%BeginExpansion
\begin{figure}[ptb]%
\centering
\includegraphics[
height=1.9675in,
width=4.8319in
]%
{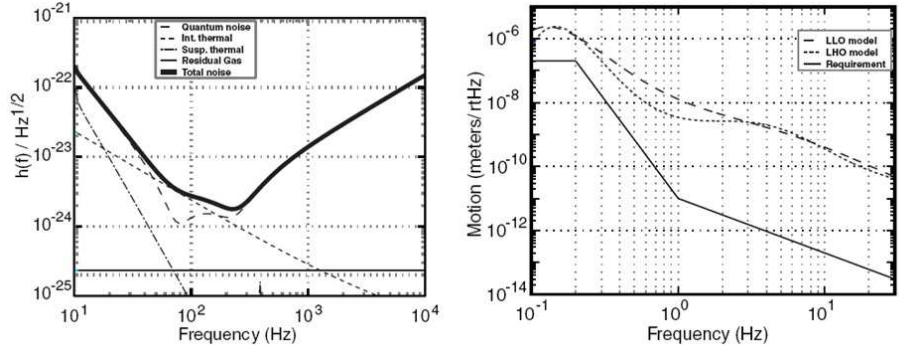}%
\caption{Left: Sketch of aLIGO projected strain sensitivity. Right:
Approximate displacement noise spectral densities at the two LIGO sites
Hanford and Livingston, along with the requirement (solid curve) on
displacement noise where the seismic isolation system supports the test mass
suspension (from ref. \cite{Abbot--rana-CQG}).}%
\end{figure}
%EndExpansion

%

%TCIMACRO{\FRAME{ftbpFU}{3.3907in}{2.6299in}{0pt}{\Qcb{Measured ground noise
%power spectral density during quiet times at a typical site in India,
%pre-selected using seismic zone maps (prepared by Rajesh Nayak and Supriyo
%Mitra, IndIGO consortium. More detailed and updated information is available
%at www.gw-indigo.org).}}{}{fig7.eps}{\special{ language "Scientific Word";
%type "GRAPHIC";  maintain-aspect-ratio TRUE;  display "USEDEF";
%valid_file "F";  width 3.3907in;  height 2.6299in;  depth 0pt;
%original-width 7.2451in;  original-height 5.6106in;  cropleft "0";
%filename 'Fig-new/fig7.eps';file-properties "XNPEU";}} }%
%BeginExpansion
\begin{figure}[ptb]%
\centering
\includegraphics[
height=2.6299in,
width=3.3907in
]%
{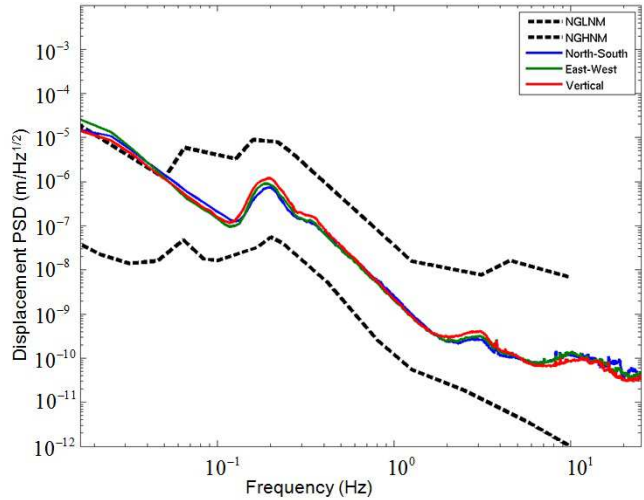}%
\caption{Measured ground noise power spectral density during quiet times at a
typical site in India, pre-selected using seismic zone maps (prepared by
Rajesh Nayak and Supriyo Mitra, IndIGO consortium. More detailed and updated
information is available at www.gw-indigo.org).}%
\end{figure}
%EndExpansion
As for development of other infrastructure, the review of technical drawings
on the interferometer stations and the UHV infrastructure has progressed well
enabling readiness for quick start after the cabinet approval. Capability for
handling the pre-stabilized laser as well as for the fabrication and welding
of the silica fiber suspensions of optics is now developed and the sense of
technical readiness is strong.

\section{The era of LIGO-India operation and GW astronomy}

India already has a strong community of young GW researchers involved in data
analysis and source modeling aspects. Therefore, the user community for
LIGO-data within India is substantial and will grow further. In fact, it was
this aspect that was the initial driving force behind IndIGO and LIGO-India.
One focus for theoretical developments will be strategies for handling
independent noise from the different detectors, especially multi-detector
coherent searches for sources where the data from different detectors in the
network is combined and analyzed with the phase information (aperture
synthesis), instead of just coincidences in time. The goal is efficient
coherent search for GW signal from binary mergers using data from global
detector network. The interface of cosmology (through CMBR) and GW is another
key area of interest here. What is envisaged is an integrated environment for
astronomy data storage and handling, more or less mirroring the evolving
vision within the GW community that GW astronomy will need simultaneous and
triggered observations by different types of telescopes and detectors,
spanning the electromagnetic spectrum, as well as arrays of cosmic ray
detectors. An advanced data centre with large computing power and storage is
integral part of the LIGO-India plan and the first stage of the centre is
being set up at IUCAA, Pune. This data centre with several hundred Terabytes
storage and about 100 Tflops peak computing power is now operational and will
eventually serve, with appropriate upgrades, as a Tier-2 data archival and
computing centre for signals from the global GW detector network.

\section{Summary}

I have sketched some key developments on the large canvas of the LIGO-India
project, the extend of which is vast, transformational for Indian science,
technology and higher education, and rejuvenating for Indian physics and astronomy.

\section{Acknowledgments}

I thank my colleagues in the IndIGO consortium and its council who contributed
much in the past 3 years to the developments described in this paper. IndIGO
has grown ten-fold by 2015, with about 120 members, working together to
realize LIGO-India and GW astronomy. Presentations on LIGO-India by Bala Iyer
and Tarun Souradeep (IndIGO council) and Stan Whitcomb (LIGO-USA) on various
occasions were of immense help in preparing this article. I thank Wei Tou Ni
and Bala Iyer for the opportunity to publish this paper based on my talk at
the Fifth ASTROD symposium (July 2012) at the Raman Research Institute,
Bangalore, India.


\begin{thebibliography}{99}                                                                                               %


\bibitem {aLIGO}G. M. Harry (for the LIGO Scientific Collaboration) Class.
Quantum Grav. \textbf{27} (2010) 084006.

\bibitem {Virgo}Advanced Virgo, https://wwwcascina.virgo.infn.it/advirgo/docs.html

\bibitem {GEO-hf}B. Willke \textit{et al}., Class. Quantum Grav. \textbf{23}
(2006) S207.

\bibitem {LISA}Laser Interferometer Space Antenna, see details at lisa.nasa.gov/

\bibitem {iseg2009}International Symposium on Experimental Gravitation, Kochi,
India (2009): see details at www.tifr.res.in/\symbol{126}iseg

\bibitem {AIGO}P. Barriga \textit{et al}, Class. Quantum Grav. \textbf{27}
(2010) 084005.

\bibitem {LIGO-Aus-DPR}LIGO-Australia - On the Crest of the Wave, The
LIGO-Australia proposal, see www.aigo.org.au/aigo\_web\_docs/LIGO-AustraliaProposal.pdf

\bibitem {GWIC-rmap}The Gravitational Wave International Committee Roadmap,
public document available at https://gwic.ligo.org/roadmap/

\bibitem {LIGO-India}B. Iyer \textit{et al}., LIGO-India Tech. rep. (2011),
LIGO document M1100296-v2, at https://dcc.ligo.org. Also available from http://www.gw-indigo.org/tiki-index.php?page=LIGO-India.

\bibitem {US-state-dept}U.S.-India Bilateral Cooperation on Science and
Technology', fact sheet at http://www.state.gov/r/pa/prs/ps/2012/06/192271.htm

\bibitem {Klimenko}S. Klimenko \textit{et al}, Phys.Rev.D \textbf{83} (2011) 102001.

\bibitem {Fairhurst}S. Fairhurst, arXiv:1205.6611v2 (gr-qc).

\bibitem {Fair-Sathya}B. S. Sathyaprakash \textit{et al.}, LIGO document
T1200219-v1, at https://dcc.ligo.org.

\bibitem {Schutz2011}B. F. Schutz, Class. Quantum Grav. \textbf{28} (2011) 125023.

\bibitem {Astrosat}B. Paul, ASTROSAT: Some Key Science Prospects, Int. J. Mod.
Phys. D22 (2013) 1341009. (The satellite was launched on 28 September, 2015).

\bibitem {hagar}HAGAR gamma ray laboratory at Hanle, Ladakh, India:
www.tifr.res.in/\symbol{126}hagar/experiment.html

\bibitem {INO}The India-based Neutrino Observatory: www.ino.tifr.res.in/

\bibitem {GMRT}The Giant Meter-wave Radio Telescope, Pune, India: gmrt.ncra.tifr.res.in/

\bibitem {Raji-unni}G. Rajalakshmi and C. S. Unnikrishnan, Class, Quantum
Grav. \textbf{27} (2010) 215007.

\bibitem {Kasevich}S. Dimopoulos, P. W. Graham, J. M. Hogan, M. A. Kasevich,
and S. Rajendran, Phys. Lett. B \textbf{678} (2009) 27.

\bibitem {Unni-gill-PLA2012}C. S. Unnikrishnan and G. T. Gillies, Phys. Lett.
A 377 (2012) 60.

\bibitem {Astrod}G. Wang and W-T. Ni, Chinese Astron. Astrophys. \textbf{36}
(2012) 211.

\bibitem {eLISA}P. Amaro-Seoane \textit{et al}., arxiv:1201.3621.

\bibitem {Decigo}S Kawamura \textit{et al.}, J. Phys.: Conf. Ser. \textbf{122}
(2008) 012006 .

\bibitem {GG}A. M. Nobili \textit{et al.}, Class. Quantum Grav. \textbf{29}
(2012) 184011.

\bibitem {Abbot--rana-CQG}R. Abbott \textit{et al}., Class. Quantum Grav.
\textbf{19} (2002) 1591.
\end{thebibliography}
\end{document}